\documentclass[runningheads]{svmult}
\usepackage{makeidx}   % allows index generation
\usepackage{graphicx}  % standard LaTeX graphics tool
\usepackage{subeqnar}  % subnumbers individual equations
\usepackage{multicol}  % used for the two-column index
\usepackage{cropmark} % cropmarks for pages without
\usepackage{physprbb}  % centered layout of diverse elements, etc.
\newcommand{\age}{\mathrel{\hbox{\rlap{\hbox{\lower4pt\hbox{$\sim$}}}\hbox{$>$}}}}

\newcommand{\eg}{e.g., } 

\newcommand{\etal}{et al.}

\newcommand{\gae}{\mathrel{>\kern-1.0em\lower0.9ex \hbox{$\sim$}}}

\newcommand{\gsim}{\!\!\!\phantom{\ge}\smash{\buildrel{}\over {\lower2.5dd\hbox{$\buildrel{\lower2dd\hbox{$\displaystyle>$}}\over \sim$}}}\,\,}
\newcommand{\gtap}{\mathrel{\hbox{\rlap{\lower.55ex \hbox {$\sim$}}\kern-.3em \raise.4ex \hbox{$>$}}}}
\newcommand{\gtrsim}{\mathrel{\hbox{\rlap{\hbox{\lower4pt\hbox{$\sim$}}}\hbox{$>$}}}}

\newcommand{\kms}{\mbox{~km s$^{-1}$}}

\newcommand{\lae}{\mathrel{<\kern-1.0em\lower0.9ex \hbox{$\sim$}}}
\newcommand{\lesssim}{\mathrel{\hbox{\rlap{\hbox{\lower4pt\hbox{$\sim$}}}\hbox{$<$}}}}
\newcommand{\lsim}{\!\!\!\phantom{\le}\smash{\buildrel{}\over {\lower2.5dd\hbox{$\buildrel{\lower2dd\hbox{$\displaystyle<$}}\over \sim$}}}\,\,}

\newcommand{\ltap}{\mathrel{\hbox{\rlap{\lower.55ex \hbox {$\sim$}} \kern-.3em \raise.4ex \hbox{$<$}}}}
\newcommand{\ltsima}{$\; \buildrel < \over \sim \;$}

\newcommand{\Msun}{\mbox{~M$_\odot$}}

\newcommand{\simlt}{\lower.5ex\hbox{\ltsima}} %< over ~

\def\E{{\sl Einstein}}
\def\ESO{{\sl European Southern Observatory (ESO)}}

\def\I{{\sl IRAS}}

\def\LMC{{\sl Large Magellanic Cloud (LMC)}}

\def\WFPC2{{\sl Wide-Field Planetary Camera 2 (WFPC2)}}
\def\WFPc2{{\sl WFPC2}}

\def\CSM{circumstellar medium (CSM)}
\def\CSm{CSM}

\def \aap #1 #2 {{Astron. Astrophys.\/} {\bf #1}, #2~}
\def \aar #1 #2 {{Astron. Astrophys. Rev.\/} {\bf #1}, #2~}
\def \aas #1 #2 {{Astron. Astrophys. Suppl. Ser.\/} {\bf #1}, #2~}
\def \aj #1 #2 {{Astron. J.\/} {\bf #1}, #2~}
\def \al #1 #2 {{Astron. Lett.\/} {\bf #1}, #2~}
\def \an #1 #2 {{Astron. Nach.\/} {\bf #1}, #2~}
\def \annap #1 #2 {{Annals Ap.\/} {\bf #1}, #2~}
\def \aph #1 {{astro-ph\/} {#1}~}
\def \ar #1 #2 {{Astron. Rep.\/} {\bf #1}, #2~}
\def \araap #1 #2 {{Ann. Rev. Astron. Astrophys.\/} {\bf #1}, #2~}
\def \asiagoc #1 #2 {{Asiago Contr.\/} {\bf #1}, #2~}
\def \apj #1 #2 {{Astrophys. J.\/} {\bf #1}, #2~}
\def \apjl #1 #2 {{Astrophys. J. Lett.\/} {\bf #1}, #2~}
\def \apjs #1 #2 {{Astrophys. J. Suppl.\/} {\bf #1}, #2~}
\def \apjsub #1 {{Astrophys. J.\/} {#1}~}
\def \apph #1 #2 {{Astropart. Phys.\/} {\bf #1}, #2~}
\def \apss #1 #2 {{Astrophys. Space Sci.\/} {\bf #1}, #2~}
\def \aspc #1 #2 {{ASP Conf.~Proc.\/} {\bf #1}, #2~}
\def \aspl #1 {{ASP Leaflet\/} {#1}~}
\def \asr #1 #2 {{Adv. Space Res.\/} {\bf #1}, #2~}
\def \astrl #1 #2 {{Astron. Lett.\/} {\bf #1}, #2~}
\def \azh #1 #2 {{Astron. Zhurnal\/} {\bf #1}, #2~}
\def \baas #1 #2 {{Bull. Am. Astron. Soc.\/} {\bf #1}, #2~}
\def \ban #1 #2 {{Bull. Astron. Inst. Neth.\/} {\bf #1}, #2~}
\def \basi #1 #2 {{Bull. Astron. Soc. India\/} {\bf #1}, #2~}
\def \ca #1 #2 {{Chinese Astron.\/} {\bf #1}, #2~}
\def \coap #1 #2 {{Contrib. Oss. Astrofis. Padova in Asiago\/} {\bf #1}, #2~}
\def \cap #1 #2 {{Comm. Astrophys.\/} {\bf #1}, #2~}
\def \emsg #1 {{ESO Messenger\/} {#1}~}
\def \gcn #1 {{GCN\/} {#1}~}
\def \hast #1 #2 {{Highlights of Astronomy\/} {\bf #1}, #2~}
\def \iauc #1 {{IAUC\/} {#1}~}
\def \iaus #1 #2 {{IAU Symp. 110: VLBI \& Compact Radio Sources\/} {\bf #1}, #2~}

\def \jcam #1 #2 {{J. Comp. Appl. Math.\/} {\bf #1}, #2~}
\def \jet #1 #2 {{JETP Lett.\/} {\bf #1}, #2~}
\def \jha #1 #2 {{J. Hist. Astron.\/} {\bf #1}, #2~}
\def \jrasc #1 #2 {{J. R. Astron. Soc. Canada\/} {\bf #1}, #2~}
\def \mem #1 #2 {{Mem. R. Astron. Soc.\/} {\bf #1}, #2~}
\def \mess #1 #2 {{The Messenger\/} {\bf #1}, #2~}
\def \mnras #1 #2 {{Mon. Not. R. Astron. Soc.\/} {\bf #1}, #2~}
\def \mplb #1 #2 {{Mod. Phys. Lett. B\/} {\bf #1}, #2~}
\def \nat #1 #2 {{Nature\/} {\bf #1}, #2~}
\def \newa #1 #2 {{New Astron.\/} {\bf #1}, #2~}
\def \nuca #1 #2 {{Nucl. Phys. A\/} {\bf #1}, #2~}
\def \nucb #1 #2 {{Nucl. Phys. B\/} {\bf #1}, #2~}
\def \npps #1 #2 {{Nucl. Phys. Proc. Suppl.\/} {\bf #1}, #2~}
\def \nyasa #1 #2 {{NY Acad. Sci. Ann.\/} {\bf #1}, #2~}
\def \obsy #1 #2 {{The Observatory\/} {\bf #1}, #2~}
\def \phfl #1 #2 {{Phys. Fluids\/} {\bf #1}, #2~}
\def \phytd #1 #2 {{Phys. Today\/} {\bf #1}, #2~}
\def \prl #1 #2 {{Phys. Rev. Lett.\/} {\bf #1}, #2~}
\def \prp #1 #2 {{Phys. Rep.\/} {\bf #1}, #2~}
\def \phyr #1 #2 {{Phys. Rev.\/} {\bf #1}, #2~}
\def \phyrd #1 #2 {{Phys. Rev. D\/} {\bf #1}, #2~}
\def \prasa #1 #2 {{Proc. Astron. Soc. Australia\/} {\bf #1}, #2~}
\def \pasa #1 #2 {{Pub. Astron. Soc. Australia\/} {\bf #1}, #2~}
\def \pasj #1 #2 {{Pub. Astron. Soc. Japan\/} {\bf #1}, #2~}
\def \pasp #1 #2 {{Pub. Astron. Soc. Pacific\/} {\bf #1}, #2~}
\def \qjras #1 #2 {{Q. J. R. Astron. Soc.\/} {\bf #1}, #2~}
\def \rma #1 #2 {{Rev. Mod. Astron.\/} {\bf #1}, #2~}
\def \rpp #1 #2 {{Rep. Prog. Phys.\/} {\bf #1}, #2~}
\def \rpph #1 #2 {{Rev. Plasma Phys.\/} {\bf #1}, #2~}
\def \sait #1 #2 {{Mem.\ Soc.\ Astron.\ It.\/} {\bf #1}, #2~}
\def \sast #1 #2 {{Sov. Astron.\/} {\bf #1}, #2~}
\def \sal #1 #2 {{Sov. Astron. Lett.\/} {\bf #1}, #2~}
\def \sat #1 #2 {{Sky \& Tel.\/} {\bf #1}, #2~}
\def \sci #1 #2 {{Science\/} {\bf #1}, #2~}
\def \spie #1 #2 {{SPIE\/} {\bf #1}, #2~}
\def \shns #1 #2 {{Stud. Hist. Nat. Sci.\/} {\bf #1}, #2~}
\def \va #1 #2 {{Vist. Astron.\/} {\bf #1}, #2~}

\begin{document}
\title*{Classification of Supernovae}
\toctitle{Classification of Supernovae}
\titlerunning{Classification of Supernovae}
% allows abbreviation of title, if the full title is too long
% to fit in the running head
%
\author{Massimo Turatto}
\authorrunning{Turatto}
\institute{Osservatorio Astronomico di Padova, vicolo dell'Osservatorio 5, I-35122, Padova, Italy}

\maketitle              % typesets the title of the contribution

\begin{abstract}
The current classification scheme for supernovae is presented.
The main observational features of the supernova types are described and
the physical implications briefly addressed.
Differences between the homogeneous thermonuclear type Ia and 
similarities among the heterogeneous 
core collapse type Ib, Ic and II are highlighted.
Transforming type IIb, narrow line type IIn, supernovae associated with 
GRBs and few peculiar objects are also discussed.

\end{abstract}

\section{Introduction} 

An adequate and satisfactory taxonomy for a class of objects should fulfill
certain elementary characteristics such as mutual exclusion,
exhaustiveness, non-ambiguity, repeatability and usefulness. The taxonomy
of supernovae (SNe) has been progressively developing since 1941 when
Minkowskii \cite{mink} first recognized that at least two main types of SNe
exist. Since then several new types of SNe have been introduced, some have
been dismissed, and others have held their ground. After an early ``Linnaeus stage''
during which the classification was based on the recognition of the
observational characteristics, the concept of SN type has been refined to
include ``genetic'' information. Nevertheless the present day classification
scheme is still not satisfactory. In particular, it is ambiguous and non-exhaustive (many objects are still classified as ``peculiar'') and the
nomenclature of the taxonomic groups, determined by historical reasons, is
confusing. However, since classification is a process which mankind
naturally and instinctively carries out in order to sort out and understand
vast arrays, the available SN classification scheme remains extremely
important and useful.

Extensive reviews on the SN taxonomy can be found in \cite{fil97,harwhe,wheben00}.  Here we give a brief update on the subject. More
details can be found in other chapters in this volume.

\section{Main Supernova Types}
The classification of SNe is generally performed on the optical spectra but,
to some extent, also on their light curves. Since SNe are brighter near their maximum light, for obvious reasons the classification is based
on the early spectra, which consist of a thermal continuum and P-Cygni
profiles of lines formed by resonant scattering. This means that the  SN
types are assigned on the basis of the chemical and physical properties of the
outermost layers of the exploding stars. Only in recent years have late time
observations  contributed to differentiating various subtypes.

\begin{figure}[t] 
\rotatebox{-0}{\includegraphics[width=.99\textwidth]{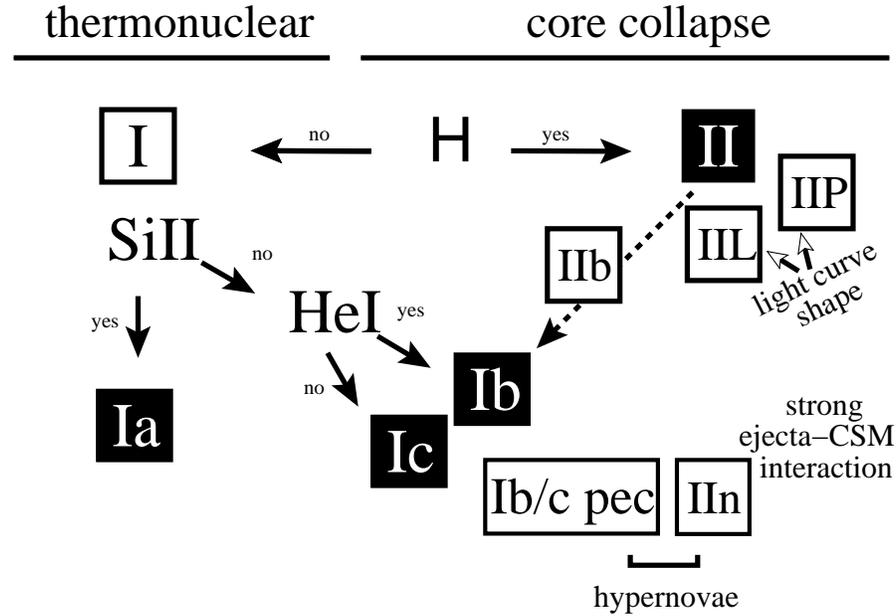}}
\caption{The current classification scheme of supernovae. 
Type Ia SNe are associated with the thermonuclear 
explosion of accreting white dwarfs. Other SN types are associated with 
the core collapse of massive stars. Some type Ib/c and IIn SNe
with explosion energies $E > 10^{52}$ erg are often called
hypernovae.}
\label{taxo} 
\end{figure}

The first two main classes of SNe were identified \cite{mink} on the basis
of the presence or absence of hydrogen lines in their spectra: SNe of type I (SNI)
did not show H lines, while those with the obvious presence of H lines were
called type II (SNII). Type I SNe were also characterized by a deep
absorption at  6150 \AA\/ which was not present in the spectra of some
objects, therefore considered peculiar \cite{bert64,bert65}. In 1965, Zwicky \cite{zwicky} introduced a 
schema of five classes but in recent years the scarcely populated types III,
IV and V have been generally included among type II SNe.

In the mid-1980s, evidence began to accumulate that the
peculiar SNI formed a class physically distinct from the others.
The objects of the new class, 
characterized by the presence of HeI \cite{gask86,hark87}, were called type
Ib (SNIb), and ``classical'' SNI were renamed as type~Ia (SNIa).
The new class further branched
into another variety, SNIc, based on the absence 
of He~I lines. Whether these are physically distinct types of objects
has been long debated \cite{harwhe,whehar}.
In several contexts they are referred to as SNIb/c.

\begin{figure}[t] 
\rotatebox{-0}{\includegraphics[width=1.05\textwidth]{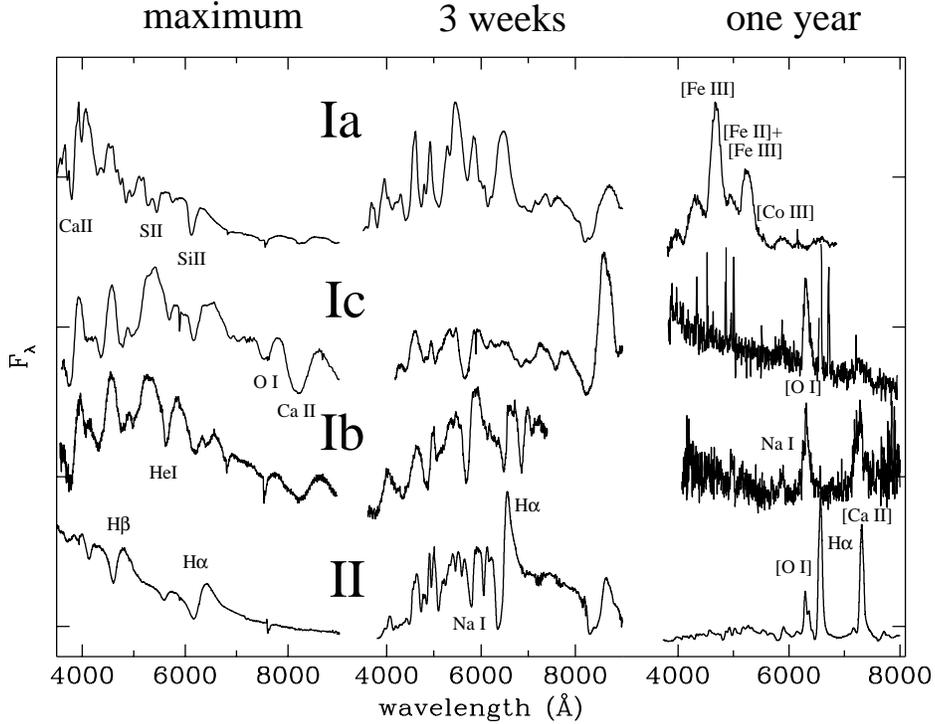}}
\caption{The spectra of the main SN types at maximum, three weeks, and one year
after maximum. The representative spectra are those of SN1996X for type Ia 
\cite{sal01}, of SN1994I (left and center) \cite{cloc96} and SN1997B 
(right) for type Ic, of SN1999dn (left and center) and SN1990I (right) for type
Ib, and of SN1987A \cite{phil} for type II. At late times (especially in the
case of the type Ic SN1997B) the contamination from the host galaxy is evident as an
underlying continuum plus unresolved emission lines. In all figures of this paper
the spectra have been transformed to the parent galaxy rest frame.}
\label{sp} 
\end{figure}

\section{Type Ia Supernovae} 

Type Ia SNe have become very popular in the last decade because of their
role in determining the geometry of the Universe with their
high luminosity and relatively small luminosity dispersion at maximum.
An extensive review of the properties of SNIa has recently appeared
\cite{leib00}.
 
SNIa are discovered in all types of galaxies, also in ellipticals 
\cite{barb99}, and are not associated with the arms of spirals as
strongly as other SN types \cite{mcmill,vdyk99}. The
spectra are characterized by lines of intermediate mass
elements such as calcium, oxygen, silicon and sulfur during the peak phase
and by the absence of H at any time (see, \eg Fig.~\ref{sp}). With age the
contribution of the Fe lines increases and 
several months past maximum the spectra are
dominated by [Fe~II] and [Fe~III] lines. The overall homogeneous spectroscopic
and photometric behavior has led to a general consensus that they are
associated with the thermonuclear explosion of a white dwarf \cite{bran95}.

Nevertheless, during the past decade early suggestions of significant differences
among SNIa \cite{barb73,psk67} have been confirmed by new, high
signal-to-noise data. The crucial year was 1991, when the bright,
slowly declining SN1991T \cite{lira98,ruiz92}, and the faint, intrinsically
red and fast declining SN1991bg \cite{fili91bg,leib93,tur91bg}
were discovered. Other under- and over-luminous objects have been found since then \cite{how01}.

The analysis of homogeneous sets of optical data led to the
discovery of a correlation between the peak luminosity and the shape of the
early light curve  with brighter objects having a slower rate of decline
than dimmer ones \cite{perl97,phil93,phil99,riess98}.
This correlation has been employed in restoring SNIa as useful distance
indicators up to cosmological distances. A correlation between the 
photometric and the spectroscopic properties was also found \cite{nugent}.

It is known that the peak luminosity of SNIa is directly linked to the
amount of radioactive $^{56}$Ni produced in the explosion \cite{arn82,arn85}.
Hence SNIa, having different magnitudes at maximum, are probably the result
of the synthesis of different amounts of radioactive $^{56}$Ni. Moreover,
there are indications of large variances (up to a factor 2) in the
total mass of the ejecta \cite{capp97}.

Observations at other wavelengths have provided very useful
information. In particular,
infrared and I-band light curves have shown that the light curves of SNIa
are characterized by a secondary peak 20-30 days after the B
maximum \cite{elias,lira98,meikle,sunt96}. Remarkable
exceptions are the faint objects like SN1991bg.

The afore mentioned diversity of SNIa persists up to very late epochs. The light
curves of faint objects are steeper than those of other SNIa, probably
because of a progressive transparency to positrons from radioactive decay
\cite{capp97,milne}. 
Faint SNIa also show slower expansion velocity of the emitting gas
both at early \cite{fil89,vdbbra} and late epochs \cite{mazz98}.

The calibration of the absolute magnitudes of a number of type Ia SNe by
Cepheids has provided the zero point for determining the Hubble
Constant (H$_0$). Average values based on limited samples and different recipes
range from M$_V=-19.34$ to M$_V=-19.64$  with small dispersions
\cite{gibson,jha99,saha,sunt99}. The recent
determination of the Cepheid distance to the host galaxy of SN1991T seems
to confirm that it is brighter (M$_V=-19.85\pm0.29$) than the 
spectroscopically normal supernovae \cite{saha01}. 

SN1991T was also the first SNIa to show photometric and spectroscopic
evidence of a light echo from circumstellar dust \cite{schmecho}. So far the only
other SNIa to show the same phenomenon was the slowly declining,
spectroscopically
normal SN1998bu \cite{capp01}. In addition to the light echo, these two SNe
suffered strong reddening, quite unusual among known, standard SNIa, suggesting
that slow decliners may be associated to younger population objects. The
fact that, on average, SNIa in late type galaxies have slower decline rates
(hence are more luminous) than SNIa in early type galaxies 
had been already
suggested  \cite{captur01,fil89,ham00,vdbpaz}.

One current issue important for the application of SNIa to cosmology is whether
they evolve with redshift. Indeed, the lower metallicity of the progenitors at
higher redshift might result in systematic differences in brightness.

\section{Type Ib and Ic Supernovae}

Type Ib and Ic appear only in spiral type galaxies
\cite{barb99,porfil} and have been associated with a parent population
of massive stars, perhaps more massive than SNII progenitors 
\cite{vdyk99}. SNIb/c exhibit relatively strong radio emission with steep
spectral indices and fast turn-on/turn-off \cite{weil01}, which is thought to arise
from the SN shock interaction with a dense circumstellar medium \cite{chev82a,chev82b}. They are, therefore, usually thought to be associated with the core collapse of
massive stars which have been stripped of their outer H (and 
possibly He) envelope. 

The introduction of this class of SNe is recent. 
As mentioned above, they were classified with other SNI until the mid-1980s, when late time observations of SNe 1983N, 1984L and 1985F
highlighted the physical differences from SNIa \cite{chug86,filsar85,gask86,hark87,whelev85}. The characterizing features
are the absence of H and Si~II lines and the presence of He~I. The excited
levels of He producing such lines are thought to be populated by fast
electrons accelerated by $\gamma$-rays from the decay of $^{56}$Ni and
$^{56} $Co \cite{hark87,lucy91}.  It was soon recognized that some
objects did not show strong He lines \cite{whee87} and the class of 
helium poor type Ic was proposed \cite{harwhe,whehar}.

In order to investigate the physical differences between these two classes,
the signatures of He were searched carefully. The He~I $\lambda$10830 line in Type Ic was first found in the spectra of  SN1994I 
\cite{fil95} with velocity as high as 17,000 \kms\/ \cite{cloc96}. It was noted that
even as little as 0.1 M$_\odot$ of helium at such high expansion velocities
implies an energy of about $3 \times 10^{51}$ erg in the outer shell alone.
Unless very high explosion energies are involved, the amount of He has to, therefore, be much smaller \cite{cloc96}. Other SNIc have shown He lines at similar (SNe~1987M and 1988L) and slower (SN1990B \cite{cloc01}) velocities. Thus the
presence of some amount of He seems to be common to several SNIc
progenitors.

Helium has been unambiguously identified also in the spectra of the recent
type Ic SN1999cq \cite{mat00}. These lines have expansion velocities much
lower than other lines, indicating that the ejecta interacts with a
dense shell of almost pure He originating from a stellar wind or mass transfer to a
companion. These findings support the idea that SNIc differ from SNIb
by the He abundance rather than by the amount of mixing of $^{56}$Ni
in the helium envelope.

\begin{figure} 
\rotatebox{-90}{\includegraphics[width=.75\textwidth]{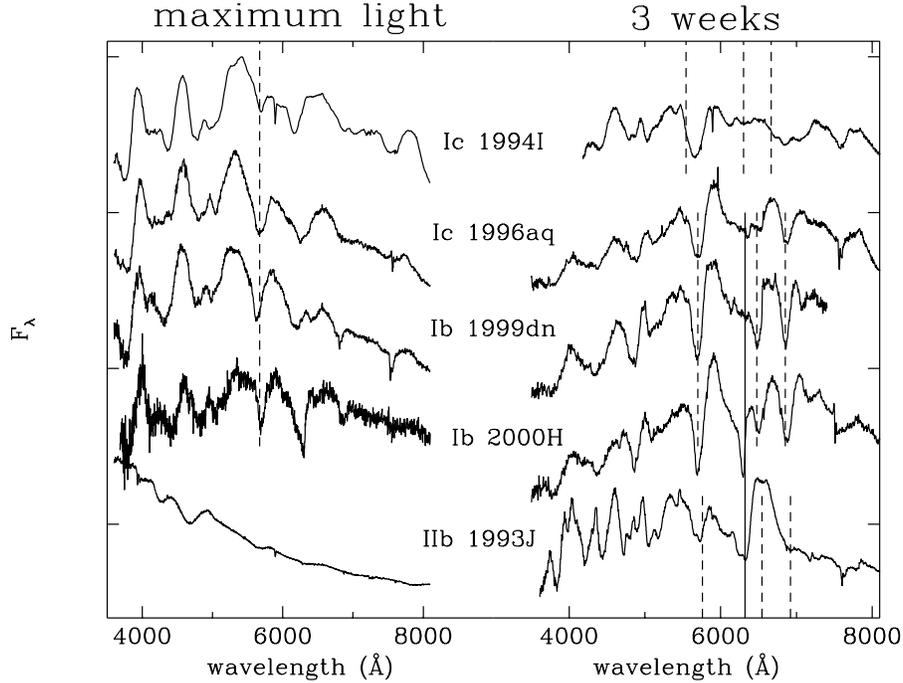}}
\caption{Comparison among the spectra of SNe reported as type Ib/c in the Asiago
SN Catalogue \cite{barb99} (left: at maximum light; right: 3 weeks
later). In addition to the spectra of SN1994I (Ic) and 1999dn (Ib) shown
in Fig.~\ref{sp},  the spectra of SN1996aq (Ic), 2000H (Ib) 
\cite{bran01} and 1993J (IIb) \cite{barb95} are displayed. With the exception of SN1993J, the
spectra at maximum are rather similar. For reference the position of He~I
$\lambda$5876 (blueshifted by 10,500 \kms) is marked with a dashed vertical line.
The dashed lines on the right indicate the positions of He~I
$\lambda$$\lambda$ 5876, 6678 and 7065 (blueshifted by 16,900 \kms\/ for 1994I,
6,000 \kms\/ for SN1993J and 9,000 \kms\/ for other objects). H$\alpha$ (solid
line, blueshifted by 11,000 \kms) is also shown. While He~I lines in SN1994I are
detectable only through a detailed analysis, they are prominent in the spectra of
SN1996aq, which should be reclassified as a SNIb.}
\label{spIb} 
\end{figure}

Absorption features attributed  to H$\alpha$ were first identified in
the spectra of the type Ib SNe~1983N and 1984L \cite{whe94}. Recent analysis
of the spectra of 11 objects has suggested that detached H is
generally present in SNIb \cite{bran01}. The optical depths of H and He are not
very high, so that modest differences in the He~I line optical depths might
transform type Ib into type Ic objects.

In addition to the different strengths of He lines, it has been suggested 
\cite{mat00} that
permitted oxygen lines are relatively stronger in type Ic than in Ib and the
nebular emission lines broader. In general, type Ib appear more
homogeneous than type Ic.

The light curves of SNIb/c have been divided in two groups depending
on the luminosity decline rate \cite{clowhe}. However, the suggestion
that type Ic include both fast and slow decliners while Ib seem to prefer slow
declines has been challenged by the existence of SNIb
with fast light curves, \eg SNe~1990I and 1991D.

With this variety, the classification of new objects based on a single
spectrum is difficult, and to some extent subjective, so that more
detailed analyses of the objects often reveal discrepancies from early
classifications. This is illustrated in Fig.~\ref{spIb} which presnets the
optical spectra of a number of objects reported as type Ib/c in the Asiago
SN Catalog \cite{barb99}.

\subsection{Type IIb} 

A few objects have been found to have early time spectra similar to
type II (i.e. with prominent H lines) and late time spectra similar to type Ib/c
SNe. For this reason they have been called  type IIb. The first SNIIb was SN1987M \cite{fil88}, but the best studied example, and one of the best studied SNe
ever, was SN1993J in M81. The analysis of seven years of observations of this
object has been recently published \cite{mat00c,mat00b}, and will
be discussed in more detail in other chapters of this volume.
More examples of transition objects are SN1996cb \cite{qiu} and
probably SN1997dd \cite{mat01}.

Fig.~\ref{spIb} shows that while the early spectrum of SN1993J was almost
featureless with a blue continuum and broad H and He~I $\lambda$5876 lines
typical of SNII, already three weeks later it displayed progressively stronger 
He~I $\lambda\lambda$ 5876, 6678 and 7065 lines characteristic of SNIb.

The light curve of SN1993J was unusual with a narrow peak followed by a secondary
maximum, recalling the behavior of SN1987A if the
time axis were reduced by a factor of four. After another rapid luminosity decline around 50 days past the
explosion, the light curve settled into an almost exponential tail with a
decline rate faster than normal SNII and similar to that of SNIa,
indicative of a small mass for the ejecta.

Indeed, the photometry of the progenitor of SN1993J taken before the explosion is
inconsistent with the spectral energy distribution of a single star, but
requires the composition of a K0Ia spectrum with a hot component
\cite{ald94}.  The radio and X-ray emission of SN1993J \cite{bart00,vand94,zimm94} have been attributed to
circumstellar interaction \cite{fran96}. Another indication comes from the
boxy shape of the emission lines of late time optical spectra \cite{mat00c,pat95}. Circumstellar gas in proximity
to the exploding star was also revealed through the detection of narrow coronal lines 
persisting for a few days after the explosion \cite{ben94}.

These SNe transforming from type II to Ib/c constitute the previously missing link
between  envelope retaining and envelope stripped SNe.

\section{Type II Supernovae} \label{snii}

Type II SNe are characterized by the obvious presence of H in their spectra.
They avoid early type galaxies \cite{barb99}, are strongly associated with
regions of recent star formation \cite{maza,vdyk99} and are commonly  associated
with the core collapse of massive stars \cite{woowea}.

SNII display a wide
variety of properties both in their light curves \cite{pat93,pat94}
and in their spectra \cite{fil97}. Four subclasses of SNII are commonly mentioned
in the literature: IIP, IIL, and IIn in addition to the above mentioned
IIb. However, a number of peculiar objects do not fit into any of these categories.

SNIIP (Plateau) and SNIIL (Linear) constitute the bulk of all SNII, and  are
often referred as normal SNII.  The subclassification is made according to the
shape of the optical light curves \cite{barb79}. The luminosity of SNIIP stops
declining shortly after maximum forming a plateau 2-3 months 
long during which a recombination wave moves through the massive hydrogen 
envelope releasing its internal
energy. SNIIL, on the other hand, show a linear,
uninterrupted luminosity decline, probably because of a lower mass 
envelope. Indeed the two classes are not separated and there are a number of
intermediate cases with short plateaus, \eg SN1992H \cite{cloc92h}. A
quantitative criterion for the classification of the light curves has been
proposed on the basis of the average decline rate of the first 100 days  
\cite{pat94}. Starting 150 days past maximum the luminosity of both types settles
into an exponential decline, consistent with complete (or constant) trapping of
the energy release of the radioactive decay of $^{56}$Co into $^{56}$Fe.

No major spectral differences exist between SNIIP and SNIIL, 
although there are recurrent claims that
SNIIL do not show the blueshifted absorption of the P-Cygni profile
evident in normal SNIIP. Indeed, a statistical analysis 
has shown that the presence or absence of the
P-Cygni absorption is correlated with the absolute brightness at maximum 
\cite{pat94}.

The progenitors of SNIIL are believed to have H envelopes of the order of
1-2 M$_\odot$, much smaller than those of SNIIP (typically 10 M$_\odot$)
probably due to mass loss during the progenitor evolution. A general
scenario has been proposed in which common envelope evolution in massive
binary systems with varying mass ratios and separations of the components
can lead to various degrees of stripping of the envelope 
\cite{nomo95}.
According to this scenario the sequence of types IIP-IIL-IIb-Ib-Ic in
Fig.~\ref{taxo} is ordered according to a decreasing mass of the envelope.

SNIIL are often radio sources \cite{weil01} and  show UV excess
attributed to Compton scattering of photospheric radiation by high speed
electrons in shock-heated \CSM\ material \cite{fran82,fran84}. A number of SNIIL, \eg SNe~1994aj and 1996L \cite{ben94aj,ben96l}, have shown lines with double P-Cygni profiles (sometimes dubbed
SNIId, ``d'' for double) indicating the presence of strong wind episodes
shortly before the explosion. These and other SNIIL show a flattening in
the light curves at late stages, a broadening of the spectral lines, and 
H$\alpha$ fluxes greater than those expected from purely radioactive models
\cite{capp95,fese93,leib91,uomo86}. These features
have been interpreted as signatures of the onset of interaction between
the ejecta and the circumstellar material.

The epochal SN1987A in the \LMC, the first SN to be observed
by naked eye in the last four centuries, was a ``not very peculiar'' type II SN
with a plateau. Extensive observations at all wavelengths have explored
the nature of this object in detail (see, \eg \cite{arn89} and references therein). 
The detection of neutrinos from SN1987A has
been a spectacular confirmation of the theory of core collapse \cite{bionta87,hira87}. The contribution of this object to our understanding of
supernovae is addressed by McCray in another chapter in this volume.

Because of its closeness, SN1987A was the first SN for which it has been
possible to unambiguously identify the progenitor, the B3~I star Sk-69 202. More recent, high
spatial resolution prediscovery images of nearby galaxies are providing new
insights into the nature of the precursor stars of other core collapse SNe.
In addition to the case of SN1993J in M81 mentioned above, tight constraints
on the mass of the progenitors of SNe~1999em and 1999gi have been published
($M({\rm 1999gi}) < 9^{+3}_{-2}$ \Msun\ and $M({\rm 1999em}) < 12\pm1$ \Msun\ \cite{sm01b,sm01a}). 

SN1987A was typical in both the explosion energy ($\sim10^{51}$
erg) and the amount of ejected radioactive material  (M$(^{56}{\rm Ni}) = 0.07$ M$_\odot$). Recent observations of other SNe seem to indicate a 
considerable dispersion
around such values. High explosion energies are required to explain the
high luminosities and kinetic energies of hypernovae (see, \eg $\S$\ref{grb})
and a large mass of radioactive $^{56}$Ni (0.3 \Msun)
has been measured for SN1992am \cite{schm94}. At the other extreme the
faint SN1997D had a considerably smaller explosion energy ($\sim {\rm few} \times 10^{50}$ erg) and  returned to the ISM a mass of radioactive material
as small as $M(^{56}{\rm Ni}) = 0.002$ \Msun\ \cite{tur97d}. The discovery
of similar objects (\eg SN1999eu) indicates that faint, under energetic
SNII may represent a
non-negligible fraction of all core collapse SNe. These
objects are the best candidates for the detection of signatures of 
black hole formation and for providing support to the 
black hole-forming SN scenario \cite{balb00,ben97d}.

\subsection{Type IIn} \label{iin}

A number of peculiar SNII have been grouped into the class of SNIIn (``n''
denoting narrow emission lines \cite{schl90}). The spectra of these objects have
a slow evolution and are dominated by strong Balmer emission lines without the
characteristic broad absorptions. The early time
continua are very blue, He~I emission is often present and, in some cases,
narrow Balmer and Na~I absorptions are visible
corresponding to expansion velocities of about 1,000 \kms
\cite{pas01},
reminiscent of SNIId (see $\S$\ref{snii}). Unresolved forbidden lines of
[OI], [OIII], and of highly ionized elements such as [FeVII], [FeX], and [AX] are
sometimes present.

One of the best cases is SN1988Z, which has been observed in the optical, radio
and X-rays for over ten years \cite{aret}. Balmer lines, in particular 
H$\alpha$, show a well defined multiple component structure with broad (FWHM
$\sim 15,000$ \kms), intermediate (2,200 \kms) and narrow 
($<700$ \kms) components with different time evolution \cite{tur88z}. 
SN1995N was a remarkably similar object, detected in radio and X-rays \cite{fox,weil01}.

It is commonly believed that what we observe in these objects is the result of interaction between the ejecta and a dense \CSm, which transforms the
mechanical energy of the ejecta  into radiation \cite{chugdan,terl92}. The interaction of the fast ejecta with the slowly expanding
\CSm\ generates a forward shock in the \CSm\ and a reverse shock in the ejecta.
The shocked material emits energetic radiation whose characteristics
strongly depend on the density of both the \CSm\ and the ejecta, and on the
properties of the shock \cite{chevfr94}. Thus the  great
diversity of observed SNIIn can provide clues to the different
history of the mass-loss in the late evolution of progenitors.

A systematic search for radio emission from SNIIn has produced contradictory
results \cite{vdyk96b}. This might be due to the high degree of
heterogeneity and to a bias in the classification of the
targets. In fact, a solid classification criterion for these SNe is
still missing, and often objects classified as SNIIn turn out to be normal
when they undergo closer scrutiny (\eg SN1989C \cite{tur93}).

A remarkable SNIIn is the recent SN1998S. Contrary to most other
SNIIn, the spectrum evolved rapidly and seems to be the result of the interaction
between the supernova and a two-component progenitor wind \cite{fass01}. 
Polarimetric studies indicate significant asphericity for its 
continuum-scattering environment \cite{leo}. 
The early spectra are well modeled by
lines arising primarily in the circumstellar region while later spectra are
dominated by the supernova ejecta \cite{lenz01}. 

\begin{figure}[t] 
\rotatebox{-90}{\includegraphics[width=.62\textwidth]{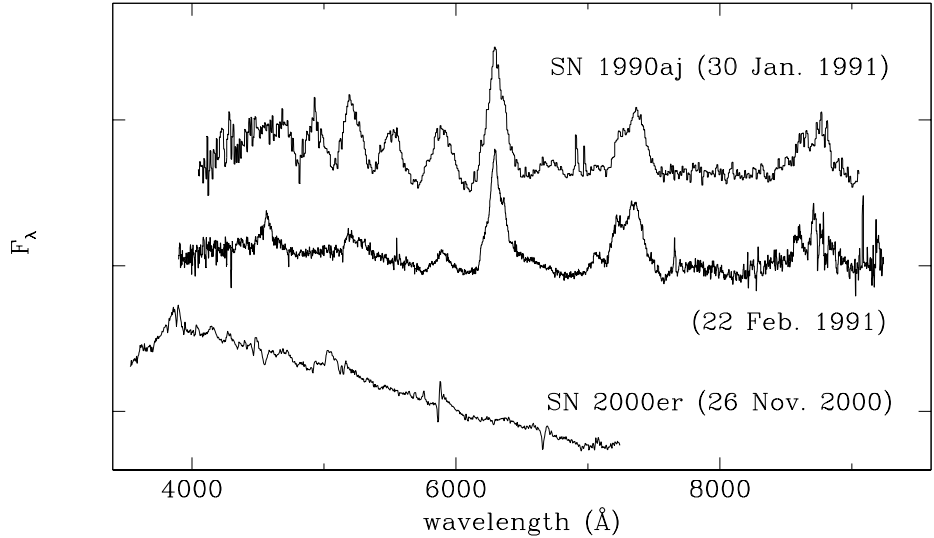}}
\caption{Spectra of SN1990aj, a SNIb/c with exceptionally
strong Fe lines at late epochs, and 
SN2000er, a bright, fast declining SN with evidence of narrow He~I lines 
in the spectra at maximum.}
\label{fig_pec} 
\end{figure}

\section{SNe and GRBs} \label{grb}

Other chapters in this volume discuss the association of SN1998bw with GRB980425 and the possibility that at least some GRBs originate from SN
explosions.

In addition to its stringent spatial and temporal association to GRB980425  
\cite{galama,iwam}, SN1998bw was peculiar in many respects. The spectrum at
early phases was unprecedented and led to different classifications 
(Ib \cite{sad98bw} or peculiar Ic \cite{fil98bw,patpiem}). 
SN1998bw was as bright as an SNIa and displayed expansion velocities as high as $3 \times 10^4$ \kms, suggesting that it was the result of an extremely energetic explosion 
($>10^{52}$ erg), even if different degrees of asymmetry and beaming allow for a
broad range of values \cite{hofl,iwam,woos98bw}. Its very powerful
radio emission has been interpreted as due to the presence of a mildly relativistic
blast wave interacting with a clumpy, structured \CSm\ deriving from a complex
mass-loss history (see the chapter by Weiler \etal\ in this volume and \cite{weil98bw}).
The complete dataset of optical and IR observations
spanning over one year after the explosion has recently appeared \cite{pat98bw}.
The analysis of the late time spectra \cite{mazz98bw} seems to confirm that the
explosion was asymmetric as suggested by polarimetry \cite{hofl}.
The presence of underlying SNe similar to SN1998bw has been invoked
to explain the anomalous rebrightening in light curves of other GRB afterglows 
at high redshift \cite{bloom,reichart}.

Other known SNe, SN1997ef \cite{garn97ef} and SN1998ey \cite{garn98ey}, bear
some spectroscopic resemblance to SN1998bw. In particular, SN1997ef was
possibly associated with GRB971115 \cite{wanwhe} and, although it was fainter
than SN1998bw, its kinetic energy was close to that of
SN1998bw \cite{mazz97ef}.
The case of the bright SNIc 1992ar \cite{cloc92ar}, 
which occurred at about two $\sigma$ from the position of 
GRB920616 \cite{woos98bw}, is also very interesting.

Additional SNe possibly associated with GRBs are SN1997cy, (GRB970514
 \cite{germ99,woos98bw}), observationally classified as SNIIn and possibly
the brightest SN ever observed \cite{tur97cy}, and its twin SN1999E 
\cite{capp99} (GRB980910 \cite{thor99}). As in the case of the type IIn
SN1988Z, they show evidence for strong ejecta-\CSm\ interaction. Explosion
energies as high as $3 \times 10^{52}$ erg are required to reproduce
the light curves \cite{tur97cy}.
 
To denote these particularly energetic objects, the 
term ``hypernova'' \cite{iwam} has been used, although its meaning is not well defined or universally accepted.

\section{Peculiar Supernovae} \label{pec}

Several objects do not fit the schema described above. In many cases they
are probably core collapse SNe exploding in unusual configurations and/or
conditions, even if the presence of other explosion mechanisms cannot be 
ruled out.

For instance, SN1993R \cite{fil97b} (and possibly SN1990aj, see, \eg Fig.~\ref{fig_pec}), showed hybrid spectral features of type Ia and Ib/c at late 
epochs. This has been interpreted as the explosion of a SNIb/c with overproduction of
$^{56}$Ni or with the slow deflagration of a white dwarf.\\ The recent SN2000er was also
classified as peculiar \cite{clotur00,maury}. The scanty 
observations available show a broad maximum at M$_V=-19$ followed by a rapid decline of 5
mag in 50 days. The spectrum shows (see, \eg Fig.~\ref{fig_pec}) a strong continuum with
a drop at wavelengths shorter than 4000 \AA, possibly due to the presence of
metals. Narrow lines  of He~I with P-Cygni profiles are visible, indicating the
presence of gas with expansion velocity $\sim900$ \kms\/ as well as
several other lines possibly due to FeII, FeIII and SiIII  \cite{bratho01}. It might be a core collapse SN which lost its He envelope shortly before the
explosion.

A well studied peculiar object is SN1961V, Zwicky's Type V, which was probably
not a genuine SN because the exploding star survived the giant eruption 
\cite{fil95b}. Similar objects might be SNe~1954J, 1999bw, and the faint SN1997bs for
which an extremely luminous supergiant precursor has been identified
\cite{vdyk99}.

\section{Conclusions} 
The taxonomy of supernovae is a subject which is still evolving. The experience of the last years has
shown that the diversity of SNe increases with our ability to detect, study,  and understand them. Even though distant SNe are disclosing new frontiers in observational
cosmology, much is still to be learned from nearby objects. Indeed, when we have the
chance to apply new observational techniques or to obtain high signal-to-noise observations of
nearby objects, new features and phenomena are revealed. Major progress 
in  understanding supernovae can be expected if
coordinated efforts involving several teams with wide ranging expertise 
and access to different observational facilities are established.
But SN occurrences are unpredictable:
in order to really understand them, we would need their cooperation.

\subsection*{Acknowledgments}
This work is partially based on data collected at the \ESO\ observatory on La Silla in Chile.

\end{document}